\documentclass[final,1p,times, sort&compress]{elsarticle}

\usepackage{amssymb}
\usepackage{comment}
\usepackage{subcaption}
\usepackage{graphicx} 
\usepackage{caption}
\usepackage{subcaption}
\usepackage{wrapfig}

\usepackage{hyperref}
\usepackage[utf8]{inputenc}

\usepackage{caption}
\usepackage{lineno}
\journal{Nuclear Physics B}

\begin{document}

\begin{frontmatter}


\title{Large-mass, low-threshold sapphire detector for rare event searches}

\author[inst1]{S. Verma}
\author[inst1]{S. Maludze}
\author[inst1]{M. Lee}
\author[inst2,inst3]{M. Chaudhuri}
\author[inst2,inst3]{V. Iyer}
\author[inst2,inst3]{V. K. S. Kashyap}
\author[inst4]{A. Kubik}
\author[inst1]{T. Lin}
\author[inst1]{R. Mahapatra}
\author[inst1]{N. Mirabolfathi}
\author[inst1]{N. Mishra}
\author[inst2,inst3]{B. Mohanty}
\author[inst1]{H. Neog}
\author[inst1]{A. Jastram}
\author[inst1]{M. Platt}

\affiliation[inst1]{organization={Department of Physics ans Astronomy, Texas A\&M University },
            addressline={578 University Dr}, 
            city={College Station},
            postcode={77840}, 
            state={TX},
            country={US}}

\affiliation[inst2]{organization={School of Physical Sciences, National Institute of Science Education and Research},
            city={Jatni},
            postcode={752050}, 
            country={India}}
            
\affiliation[inst3]{organization={Homi Bhabha National Institute},
            addressline={Training School Complex, Anushaktinagar},
            city={Mumbai},
            postcode={400094},
            country={India}}
            
\affiliation[inst4]{organization={SNOLAB},
            addressline={Creighton Mine \#9, 1039 Regional Road 24}, 
            city={Sudbury},
            postcode={ON P3Y 1N2}, 
            country={Canada}}

\begin{abstract}
Low mass nuclear recoil dark matter and coherent-elastic-neutrino-nucleus-scattering (CE$\nu$NS) searches confront similar challenges in choosing ultra low threshold and large-mass detectors. We report experimental results from the first-of-its-kind 100 g single-crystal sapphire detector design with diameter of 76 mm and thickness of 4 mm. The detector is designed to be sensitive for low-energy rare interactions with an intention to investigate the low mass region of dark matter phase-space and search for CE$\nu$NS at the reactor site. Sapphire is a crystal of aluminum oxide ($\mathrm{Al_{2}O_{3}}$) and has been found to be a good candidate for light mass spin-dependent dark matter search experiments due to its lower atomic mass compared to other detector materials such as germanium and silicon. Using the data collected from the test facility at Texas A\&M University, we were able to resolve low energy lines from calibration sources and estimated that our newly developed sapphire detector has a baseline recoil energy resolution of 18 eV. These detectors are operated at 0 V with the phonon-assisted detection providing a quenching-free low-threshold operation.


\end{abstract}




\end{frontmatter}


\section{Introduction}
Numerous astronomical observations have indicated that the majority of matter content of the Universe is composed of non-luminous, non-baryonic form of matter called ``dark matter'' \cite{Zwicky1933,Rubin1,DMcandidates}, whose particle nature remains a mystery. Various experimental constraints have been published on weakly interacting massive particle (WIMP) dark matter for masses above 10 $\mathrm{GeV/c^{2}}$ with no success in finding the particle nature of dark matter \cite{PhysRevLett.121.111302, PhysRevLett.118.021303, PhysRevD.98.102006, PhysRevD.102.091101}. Most of these searches are based on direct detection techniques \cite{Essig:2022dfa} in which dark matter elastically scatters off the atomic nucleus. The resulting nuclear recoil produces various forms of signals such as phonons, ionization, light etc. Absence of positive laboratory signal for GeV scale dark matter strongly compels experimentalists to look for low-mass dark matter. In order to investigate lower mass region of the parameter space beyond the current detection limits, it is essential to have a detector with low threshold for nuclear recoils as well as a large mass to account for the extremely low WIMP-nuclear interaction cross section.


Another process that requires low threshold detectors is coherent-elastic-neutrino-nucleus-scattering (CE$\nu$NS) which was theorized in 1974 \cite{PhysRevD.9.1389} and was first detected by COHERENT collaboration in 2017 using neutrinos from a neutron spallation source \cite{COHERENT, Arguelles:2022xxa}. In this process, a neutrino coherently scatters off an entire atomic nucleus giving a neutron number squared boost to the cross section. Since the deposited nuclear recoil energies is of the order of $\sim$ keV \cite{Abdullah:2022zue}, it faces similar challenges that of low mass dark matter detection and hence require low threshold detectors. 

In this paper we describe the fabrication procedure and working principle of a novel low threshold phonon - mediated 100 g single-crystal sapphire detector instrumented with Transition Edge Sensors (TES) and present detector characterization results from the experimental run at Texas A\&M cryogenic test facility. This new low-threshold sapphire detector is operated at 0 V with the phonon-assisted detection providing a quenching-free low-threshold operation. Lower atomic mass of Al and O in sapphire makes it sensitive to lower nuclear recoil energies thereby making it a good candidate for both dark matter and CE$\nu$NS experiments. 

\section{Detector Technology}
The detector used in this work is made up of 100 g single-crystal sapphire. It is cylindrical in shape and oriented in A-plane (11-20). It is equipped with an array of tungsten Transition Edge Sensors (TES) \cite{doi:10.1063/1.1146105} photolithographically fabricated in parallel on one side of the crystal to measure athermal phonons from lattice interactions. During the experimental run, TES sensors are voltage biased and are operating in the superconducting-normal transition region. The sensors are readout through 4 channels - inner three channels and one outer circular channel, as shown in Fig.~\ref{fig:sapphire_picture}, to facilitate position reconstruction using energy and timing. 

\subsection{Detector Fabrication}
Detector fabrication begins with high purity sapphire single crystal substrate and uses photolithography to pattern desired circuit structures on the substrate using the standard etching techniques \cite{JASTRAM201514}. The substrate is first shaped and aligned to a specific crystal axis and orientation. It is then chemically etched to remove any contaminants present on the surface due to  previous processes or exposure to air born radon contamination. Substrate surface is polished before putting any sensors on them. Before the polished substrate can be processed into detectors, thorough cleaning of the substrate is performed using piranha clean process which consists of 30\%  $\mathrm{H_{2}O_{2}}$, 98\% $\mathrm{H_{2}SO_{4}}$, 50\% HF and 37\% HCL \cite{JASTRAM201514} which removes surface contaminants as well as any particulates that may cause defects in subsequent processing. For this reason, the cleaning is performed in a class 100 clean room.

Thin films of metals that form the final circuit on the detector are deposited using a fully automated DC magnetron sputtering system with capability to deposit eight crystals simultaneously. The two thin metallic films on the detector are 600 nm of Aluminum (Al) which is an absorbing layer for the collection of phonons (Al is superconducting at the operating temperatures) and 40 nm of Tungsten (W) which is the main component of the Transition Edge Sensors (TES). The photolithography process involves spin coating the crystal with a positive photoresist which is a light-sensitive chemical, pre-baking the photoresist, exposing under a UV mask aligner with the right circuit pattern, developing the photoresist, then hard-bake followed by finally using the standard wet etching techniques which also works for Si/Ge single crystals to print required TES sensors on the crystal surface (30\% $\mathrm{H_{2}O_{2}}$ is used to etch the tungsten and Cyantek Al-11 etchant is used to etch the aluminum). 

\subsection{Working Principle}
Particle interaction in the bulk of the detector causes the atomic nucleus to recoil which generates phonons and scintillating photon signals. In this work we primarily focus on phonons which are the true measure of the recoil energy as it does not depend on the type of particle interaction. Phonons which are generated in the bulk of the detector travel to the surface and enter the absorbing aluminum layer. In aluminum which is superconducting at cryogenic temperatures, if the phonon energy is greater than twice the band gap energy of Al, they break Al cooper pairs which undergo random walk and are finally trapped into tungsten due to lower band gap energy of W compared to Al. The temperature of tungsten which is the main element of the TES is maintained near its critical temperature (between the normal and the superconducting state). Phonons absorbed by the Al fins will break cooper pairs, creating quasiparticles that are trapped in tungsten, changing its resistance. The phonon signal is read out using a SQUID front-end amplifier \cite{doi:10.1063/1.1146105}. One major difference in fabrication of sapphire detector is the absence of amorphous-Si layer which is present in CDMS style detectors to protect the substrate surface from aluminum/tungsten etchants. Aluminum (Al) is deposited directly on sapphire ($\mathrm{Al_{2}O_{3}}$) which results in better lattice matching between the substrate and absorbing Al layer and improves phonon transmission from sapphire to Al \cite{KNAPEN2018386, PhysRevD.98.115034, PhysRevD.101.055004}. 

\section{Analysis and results}
Experimental setup was mounted inside the BlueFors pulse-tube-based ${^3}He$-$^4He$ dilution refrigerator located at Texas A\&M Cryogenic Detector Testing facility, reaching a base temperature of $\sim$8 mK. Bias voltage value across the TES sensors was tuned so that tungsten reaches desired transition temperature. At that temperature we expect to have a linear response in signal amplitude up to the energy that brings tungsten to its normal conducting state. Part of the analysis goal was to examine the linearity of the detector response and choose a calibration line within the linear region. For that purpose $^{241}Am$ isotope was placed on the bare surface of the detector facing detector channel D. The source itself was placed inside a specifically designed copper source holder with a narrow slit. By optimizing the width of the slit in the copper holder, event rate from the source is reduced to desired value. In addition, particles passing through the slit are collimated, which localizes interactions to a narrow region of the detector. The $^{241}Am$ source is characterized by multiple energy lines such as 59.54 keV and 26.30 keV  $\gamma$'s from nuclear decay and $L_l$, $L_\alpha$, $L_\beta$, $L_\gamma$ characteristic X-rays from its decay product \cite{Am241_1, Am241_2}. The multiplicity of energy lines allowed us to study the dependence of detector energy resolution on different energies. The main goal of the study is to estimate the baseline resolution of our detector, i.e. the energy resolution when there are no particle interactions within the detector and the signals recorded consist primarily of electronic noise from the circuit. The baseline resolution is directly related to detector threshold, which plays a crucial role in rare event searches.  

\begin{figure}[h]
    \centering
    \includegraphics[width = 8.5cm]{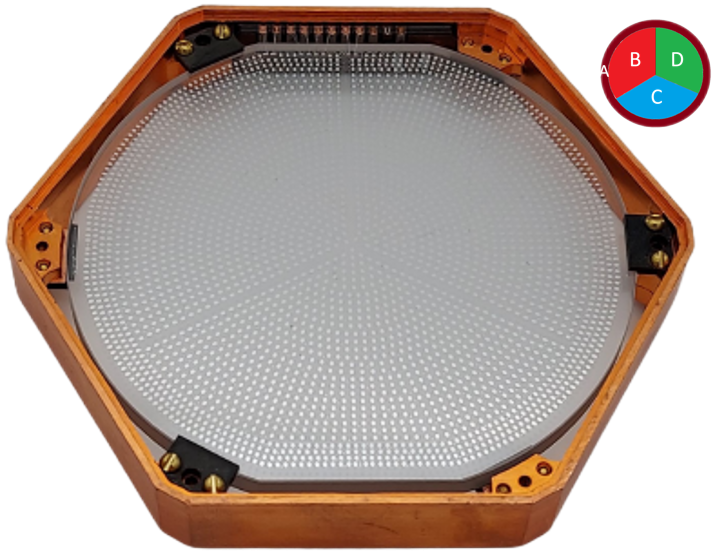}
    \caption{Top: Sapphire detector in copper housing with the crystal diameter of 76 mm and thickness of 4 mm. Reflection of the fabricated phonon sensors can be seen due to the transparency of Sapphire crystal. It is equipped with 3 inner channels which provide fiducialization for event reconstruction in the bulk of the crystal and  an outer channel to reject events near the edge \cite{PhysRevD.95.082002,  PhysRevLett.112.241302}.}
    \label{fig:sapphire_picture}
\end{figure}
\begin{figure}
\centering
\includegraphics[height = 3.8cm]{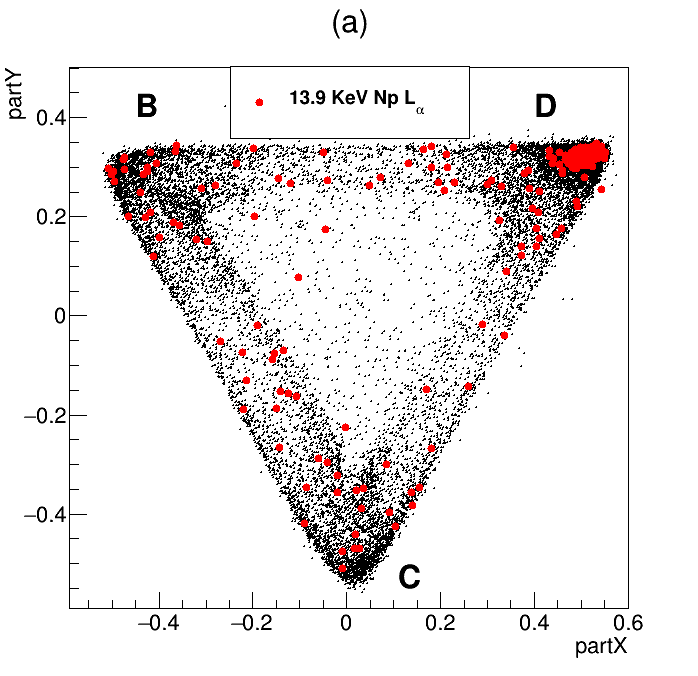}
\includegraphics[height = 3.8cm]{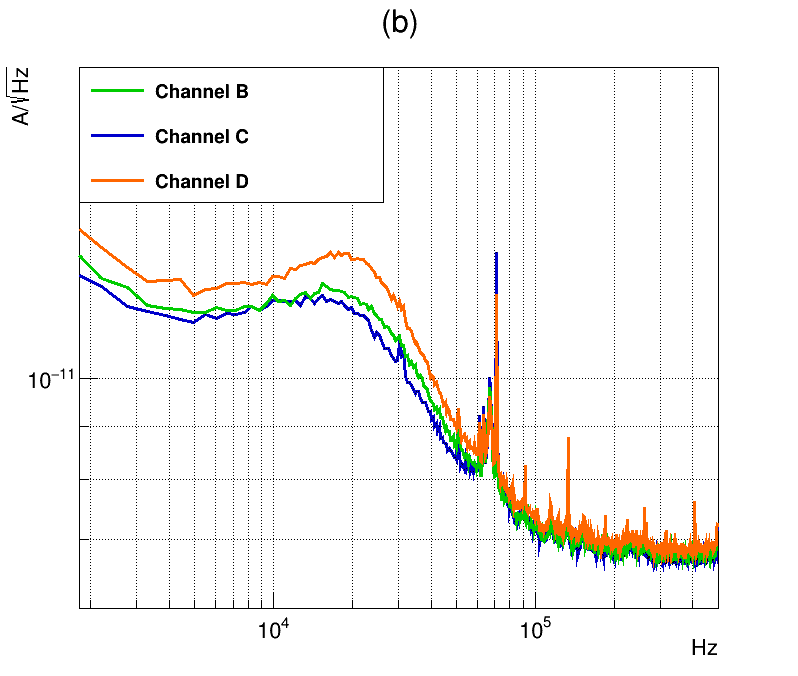}
\caption{(a) 13.9 keV X-ray (red dots) events on the partition plot. All events in range 6-70 keV are displayed in black dots. Partition variables are combinations of inner channel pulse amplitudes such that events happening under the particular channel would show up at the corresponding corner of that partition triangle. Upper right corner corresponds to Channel D where the source was placed. (b) Noise power spectral density of inner 3 channels after pulse rejection.}
\label{fig:Sapphire_partition}
\end{figure}

We performed multiple experimental runs until optimal operation conditions was reached. The run with the best performance contains approximately 3 hours of data. During the experiment, we record raw pulses - 2ms long signal (waveform) from the TES sensors (Fig. \ref{fig:Sapphire_vs_Ge}). Each pulse has approximately 2000 sample points and represents an event in the detector. First we record several minutes of randomly selected waveforms. Since the record rate is much higher than the expected event rate from the source, this first dataset contained mostly noise. Small fraction of events from particle interactions are filtered out from the analysis by rejecting high amplitude and high standard deviation waveforms. To characterize noise environment we display average noise power spectral density (average noise in frequency domain shown in Fig. \ref{fig:Sapphire_partition}b). Events with amplitude greater than a set threshold are recorded for further analysis. Trigger threshold is set in a way that recorded event rate is comparable to the rate of the calibration source. The algorithm called Optimal Filter (OF) \cite{Golwala} is used to extract amplitude of a signal event from the noisy raw pulse. The algorithm fits pulse and noise templates to the raw pulse by minimizing $\chi2$ in frequency domain. As it was mentioned before, average noise template can be found from the noise dataset by applying certain conditions to reject few pulses in the set. Similarly, pulse template is found from the main triggered dataset as an average of raw pulses. In this case, raw pulse selection criteria for template depends on the standard deviation and maximum of the flat region of a pulse. The amplitude obtained from the algorithm for each raw pulse is a direct measure of energy.  

\begin{figure}
    \centering
    \includegraphics[width = 9.5cm]{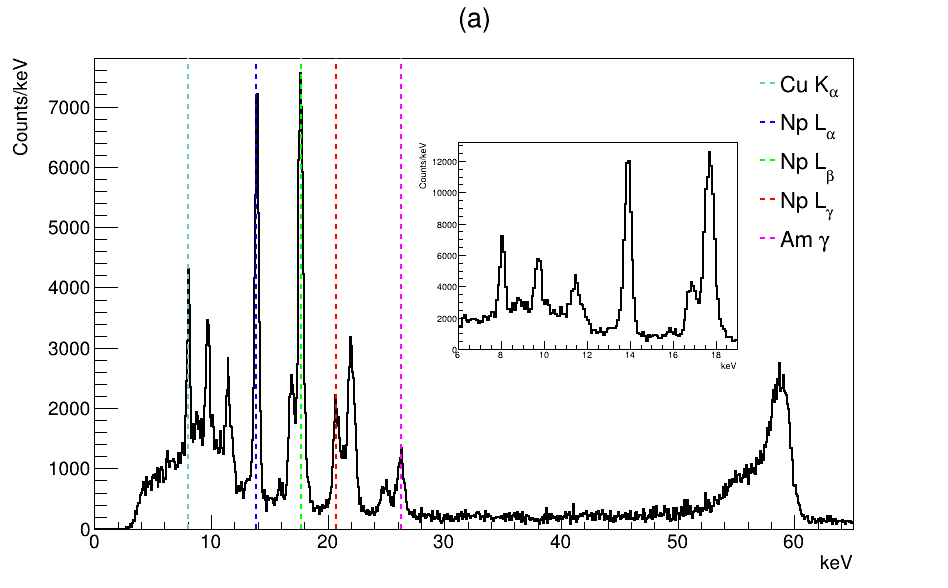}
    \includegraphics[width =9.5cm]{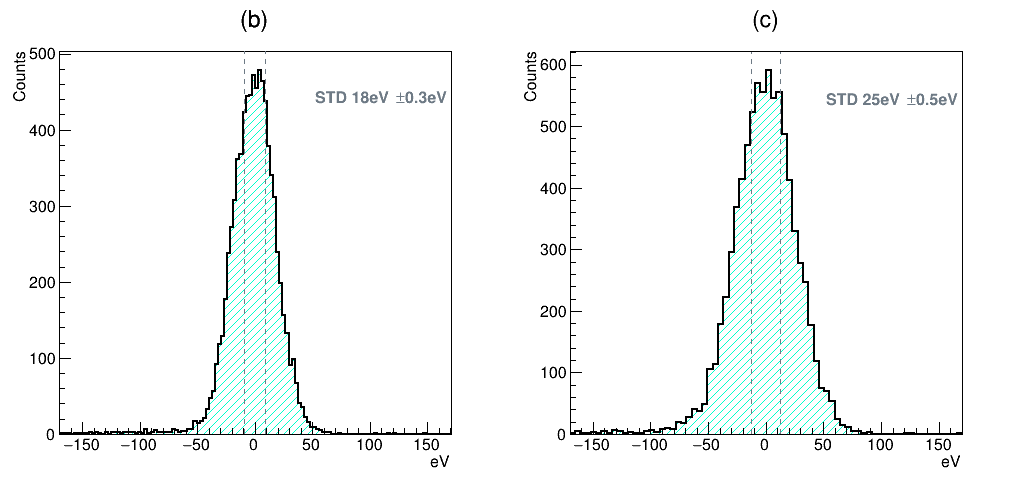}
    \caption{(a) Deposited energy distribution on the sapphire detector. The spectrum is calibrated based on 13.9 keV Np $L_\alpha$ (blue). Other dotted vertical lines correspond to X-rays from $^{241}Am$ decay product $^{239}Np$ - 17.7 keV (green), 20.7 keV (red). 26.3 keV $\gamma$ from $^{241}Am$ is indicated by vertical dotted line in magenta. 8.02 keV $K_\alpha$ X-ray line from copper is indicated by dotted line in cyan. (b) Calibrated noise amplitude distribution for channel D (Where calibration source was placed) in energy unit. The standard deviation of the Gaussian fit is 18 eV $\pm$ 0.3 eV. (c) Calibrated total (Sum of all channels) noise amplitude distribution in energy unit. Standard deviation of a Gaussian fit over the distribution is 25 eV $\pm$ 0.5 eV. 
    Dotted vertical lines represent $\mu \pm \sigma/2$ of the Gaussian fit.}
    \label{fig:Sapphire_main}
\end{figure}

\subsection{Calibration}
Calibration source $^{241}Am$ undergoes $\alpha$ decay and produces $^{239}Np$ and 59.5 keV $\gamma$'s as a byproduct. Excited state of $^{239}Np$ produces number of  characteristic X-rays, from which 13.9~keV ($L_{\alpha}$), 17.7 keV ($L_{\beta}$) and 20.7 keV $L_{\gamma}$ are the most prominent. Lower energy 13.9~keV line was chosen for calibration because TES sensors lose power to extract accurate energy information for higher energy interactions. We call this effect saturation, when higher deposited energy causes tungsten to transition to normal conducting state and amplitude of a signal is no longer proportional to the deposited energy.

Reconstructed energy spectrum based on $^{241}Am$ 13.9 keV calibration line is displayed on Fig. \ref{fig:Sapphire_main}a. As a result we have identified 11.8 keV ($L_{l}$), 13.9 keV ($L_{\alpha}$), 16.9 keV ($L_{\beta2}$), 17.7 keV ($L_{\beta1}$), 20.7 keV ($L_{\gamma}$) X-ray lines from $^{239}Np$ as well as 26.3 keV and 59.5 keV $\gamma$'s from the nuclear decay of $^{241}Am$. In addition we observed 8.05 keV, 8.91 keV Cu $K_\alpha$ and Cu $K_\beta$ energy lines from the copper source holder. Copper surrounding the crystal (Fig. \ref{fig:sapphire_picture}) can provide X-ray lines as well, but its contribution is expected to be lower since the events coming from the source are collimated and parallel to the copper housing. We can see that low energy peaks in the distribution lay close to expected energies characteristic to $^{241}Am$ isotope and copper (dotted lines on the Fig. \ref{fig:Sapphire_main}a), which indicates that the detector response is linear up to 30 keV. We observe small saturation effect on 59.5 keV events as its peak falls on a slightly lower number.
Detector inner channel layout allows us to determine relative location of an event inside the crystal by combining pulse amplitudes. Motivation behind partition variables (partX and partY) is that events happening closer to a specific channel would have higher amplitude on that channel. Characteristic triangular-shaped plot is displayed on Fig. \ref{fig:Sapphire_partition}a, where we can see that events from the source are well localized under the detector channel D. Even though the detector is perfectly capable of resolving much lower energies than 8 keV, only events with energies above the threshold were recorded for calibration purposes in this run. 
\subsection{Baseline Resolution}
Due to the electronic noise, detector response when there is no interaction is still non-zero. The electronic noise limits detector's ability to measure energy interactions below some threshold and affects resolution for higher energies. To estimate baseline resolution we use the same OF algorithm again, but this time on the noise dataset. Obtained noise amplitude distribution is calibrated in energy scale using the same factor. In Fig. \ref{fig:Sapphire_main}c, distribution of a sum of all channel amplitudes in energy scale is displayed. Baseline resolution is estimated as $1\sigma$ of the Gaussian fit to the distribution which comes out to be 25 keV. We also estimate baseline resolution for just one channel where the source was placed and the signal is the most  prominent. Knowing single channel baseline performance is important for low energy interactions, since most of the energy is expected to be deposited to the closest inner channel. Fig. \ref{fig:Sapphire_main}b displays noise distribution for channel D in energy scale where the baseline resolution was estimated to be 18 keV. 

\begin{table}[h]
\centering
\resizebox{\columnwidth}{!}{%
\begin{tabular}{|lllll|}
\hline
\textbf{Label}  & \textbf{Energy (keV)} & \textbf{Resolution (keV)} & \textbf{Error (keV)} & \textbf{Rate (Hz)} \\ \hline
$Cu K_{\alpha}$ & 8.02                 & 0.1565                   & 0.0032              & 0.14                 \\
$Np L_{\alpha}$ & 13.9                 & 0.1882                   & 0.0011              & 0.54                 \\
$Np L_{\beta}$  & 17.7                 & 0.2361                    & 0.0015              & 0.79                 \\
$Am \gamma$     & 26.3                 & 0.4306                    & 0.0100              & 0.18  \\ 
\hline
\end{tabular}
}

\caption{\label{table2} Energy resolution and event rate for different peaks observed in the distribution.}
\end{table}

\begin{figure}[h]
    \centering
    \includegraphics[width = 7cm]{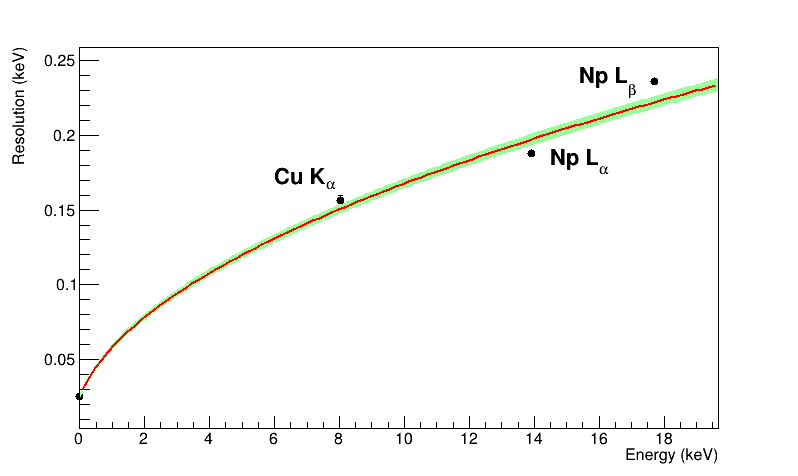}
    \caption{Energy resolution for each peak in the distribution defined as standard deviation of a Gaussian fit. Three points  on the plot correspond to Cu $L_{\alpha}$ and $^{241}Am$ lower energy lines. First point represents to baseline resolution. Red line shows the fit function and  green shaded area represent uncertainty of the fit.}
    \label{fig:Sapphire_Resolution}
\end{figure}

\subsection{Energy resolution}
In addition, having multiple energy lines observed in the spectrum enables us to estimate resolution at different energies. A simple model that describes resolution dependence on energy is a combination of a baseline noise and systematic uncertainty (deviation from the mean). Since systematic uncertainty scales as $\sqrt{E}$, function $f(x) = \sqrt{aE+{\sigma_b}^2}$ was chosen to describe the dependence, where $\sigma_b$ was fixed on the baseline resolution. In this study we chose prominent peaks at 8.05 keV, 13.9 keV and 17.7 from the distribution shown in Fig. \ref{fig:Sapphire_main}a. Fig. \ref{fig:Sapphire_Resolution} displays the fit over these three points starting from the baseline resolution value. The energy resolution values and their uncertainties for different peaks are given in Table 1.
\subsection{pulse timing}
\begin{figure}[h]
    \centering
    \includegraphics[width = 6cm]{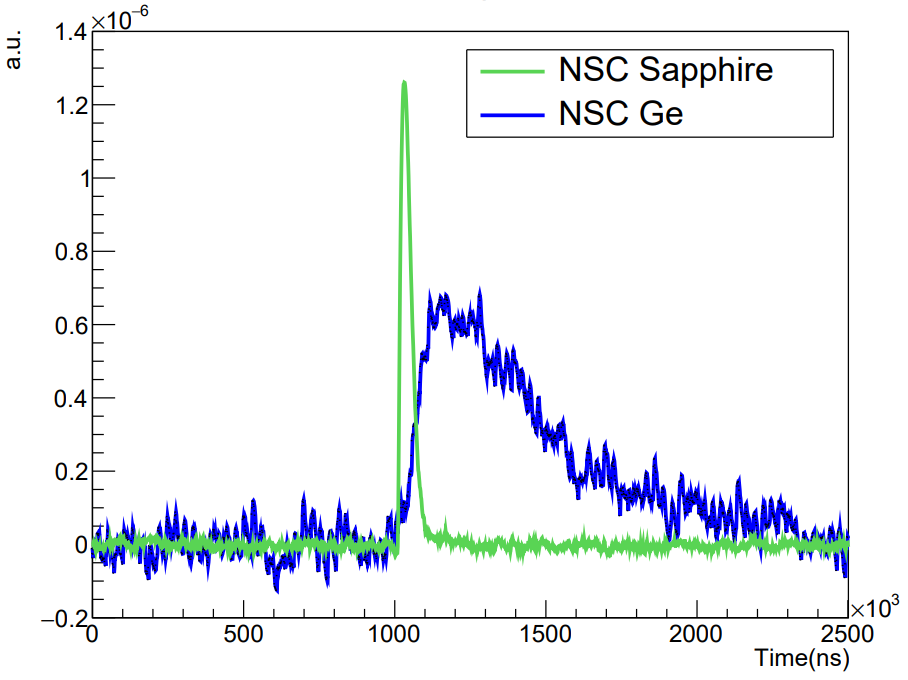}
    \caption{Pulses from similar size Germanium and Sapphire detectors within the same energy range. Pulses are selected from different experimental runs for comparison. Note that the run with the germanium detector had a higher level of noise.}
    \label{fig:Sapphire_vs_Ge}
\end{figure}

In addition we observe faster rise time of a pulse compared to other detectors recently tested and use for experiment. Fig. \ref{fig:Sapphire_vs_Ge} shows that pulses for 4mm sapphire are faster compared to similar size Germanium detector.  

\section{Summary and Outlook}
We have demonstrated 18 eV baseline resolution for one channel and 25 eV for all channels of a 100 g sapphire detector and observed distinct calibration lines from the source ($^{241}Am$ isotope), which points to the detector having excellent resolution. Another advantage of this technology lays in atomic composition of the sapphire crystal as lower atomic mass of Al and O would translate into higher recoil energies for WIMPs with comparable masses. With this low threshold and sufficiently large detector mass, this detector technology could allow us to study unexplored parameter-space for low mass dark matter search. Future R\&D efforts will be dedicated to both the simultaneous collection of light using a separate cryogenic detector, as well as pulse-shape analysis to distinguish between Electron Recoil (ER) and Nuclear Recoil (NR) events for effective background rejection. The detector is already being used in current experimental runs at Texas A\&M Nuclear Science Center as part of MINER (Mitchell Institute Neutrino Experiment at Reactor) \cite{AGNOLET201753} and a follow up paper will discuss the performance of sapphire detector in the MINER experimental setup.


\section{Acknowledgement}
This work was supported by the DOE Grant Nos DE-SC0020097, DE-SC0018981, DE-SC0017859, and DE-SC0021051. We acknowledge the seed funding provided by the Mitchell Institute for the early conceptual and prototype development. We would like to acknowledge the support of DAE-India through the project Research in Basic Sciences - Dark Matter and SERB-DST-India through the J. C. Bose Fellowship.


\bibliographystyle{elsarticle-num}


\bibliography{sapphire-ref}{}


\end{document}